# A Novel Pair and Matching Algorithm for Embedding Secret Messages in Images

**P N Priya[1], R Ranjitha[2], Yashaswini Naik[3], Shrilekha[4], Rama Moorthy H[5]**

[1,2,3,4]Computer Science Engineering, SMVITM, Bantakal, Udupi-574115
[5]Assistant Professor, CSE, SMVITM, Bantakal, Udupi-574115

**Abstract**—*Steganography has proven to be one of the practical way of securing data. It is a new kind of secret communication used to hide secret data inside other innocent digital mediums. There are various algorithms for pair and matching technique. One such method uses two bits of secret message to be matched with cover image bits. Algorithm had used only 2 pairs to be mapped [1]. Thus limiting the matching to 6 bits per pixel. Here in our proposed algorithm we are matching 3 bits at a time. Thus in the best case scenario we can match up to 9 bits per pixel.  It is also a good foundation to build more secure communication in today's data centric world.*

## I. INTRODUCTION

Today steganography has gained huge importance because of increasing use of internet by people and other new technologies. National, corporate, personal information security is some of the important concerns [5]. It is necessary to employ effective methods for confining the access to the data over any communication channel. Cryptography is called as hidden writing which deals with encrypting the data into cipher form and transmitting the data. So even if the attacker gets the cipher, data is secure unless the information about the cryptosystem used is available to him. But the major drawback is the cipher text transferred draws the attention of the eavesdropper so that he can easily identify the transmission of confidential data and cryptanalysis could be carried out with many attempts. Steganography overcomes this by hiding the data in a cover media.

The covered data is hardly noticed by eavesdropper [6] i.e. sending encoded message may draw attention but invisible information will not. Steganography is the practice of concealing a file, message, image, or video within another file, messages, images, or video. The different types of file formats used in steganography are text, image, audio, and video and protocol steganography. In image steganography spatial domain technique is mainly used which directly change image pixel values. Least Significant Bit insertion is the simplest and popular algorithm used in this technique.

## II. LITERATURE REVIEW

The Pair and Matching technique [1] uses images which contain number of pixel and each pixel contains 3 bytes

(i.e. Red, Green and Blue). These images' binary value and hidden text's binary value are paired. Now each pair of message bits are matched with the pair of image bits. If there is any matching occurs then the position value is stored in the index as shown in figure. The position values/index values are stored in the image.

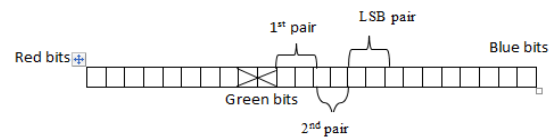

| Index 1 | Index 2 |
|---------|---------|
| 0       | 0       |
| 0       | 1       |
| 1       | 1       |
| 1       | 0       |

The flow starts by taking the cover image and the hidden text as the input. The size of cover image should be greater than the size of hidden text. The hidden text and the pixels of the image are encoded to binary and the bits are paired. Now the hidden text bits are compared with the bits of pixel. If there is no match then hidden text bits are saved in LSB bit pair and value is saved as 0 in index1 and index2. If the paired bits of hidden text are matched with first pair after MSB bit i.e. 3rd and 4th bits in image pixel position, then index value of index1 is 1 index2 is 0. If the pairing happens to be in second pair i.e. 5th and 6th bits in image pixel position, then index value of index1 is 1 index2 is 1. The process goes on until all the paired bits of hidden text are embedded. Here the disadvantage is that only 2 bits are compared.

The quality of the stego image is measured by calculating Peak to Signal Noise Ratio(PSNR).It is define as the ratio between the maximum possible power of a signal and the power of corrupting noise that effects the fidelity of its representation. It is usually defined using Mean Square Error (MSE). The formulae to calculate the MSE and PSNR for two m*n monochrome images I and K are:

$$MSE = \frac{1}{mn}\sum_{i=0}^{m-1}\sum_{j=0}^{n-1}[I(i,j) - K(i,j)]^2$$

$$PSNR = 10.\log_{10}(\frac{MAX_I^2}{MSE}) = 20.\log_{10}(\frac{MAX_I}{\sqrt{MSE}})$$

Larger the PSNR value, lower the distortion and smaller the possibility of visual attack.

## III.  PROPOSED ALGORITHMS

### A. Preprocessing

An image is selected as the cover media and the secret text is converted to binary and embedded in the image using embedding algorithm [7][8]. Then the stego image is sent in the insecure transmission medium. Following figure represents the overall process of image steganography:

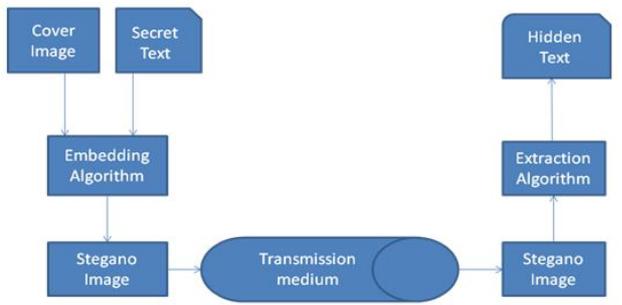

### B. Pair and Match Technique

The binary of the cover image and the secret text is taken as the input. The binary of the message bit is grouped into 3 bits. The bits, starting from the second MSB bit to sixth bit, of each byte of the image pixel is grouped into 3 bits as shown in the following figure. The groups in the pixel and the text are paired to find the match. The position of the match is inserted in the lower 2 bits of the pixel byte.

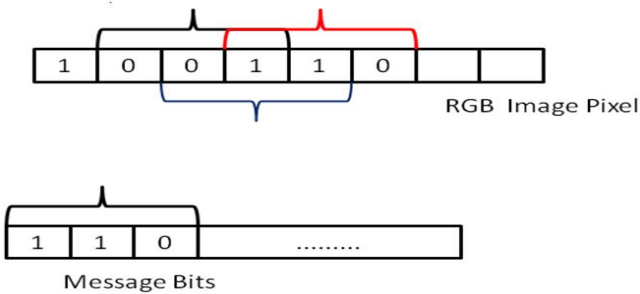

### C. Hiding Technique:

The cover image and the secret text are converted into binary values and fed as input to the embedding algorithm. The bits in pixel as well as in the text are grouped into 3 bits. The following are the steps to be carried out in the proposed embedding algorithm.

Step 1: Start

Step 2: Select a RGB cover image and the secret text to be embedded.

Step 3: Encode the cover image and hidden text into binary.

Step 4: Choose a pixel and divide it into 3 bytes(color component). Select a byte and group the bit to form 3 bit group.

Step 5: Group the secret bits into groups of 3 bits.

Step 6: Group the bits into 3 bits starting from second MSB bit to the sixth bit of pixel byte. This forms 3 groups per pixel byte.

Step 7: Compare a group of the secret bit against each group of the pixel byte.

Step 8: If match is found with the first group, 01 is inserted; Else if match is found with second group, 10 is inserted; Else if match is found with third group, insert 11 to the LSB bits of the pixel byte; Else  goto Step 10.

Step 9:Advance to the next group in the secret text. Goto Step 11

Step 10: Insert 00 to the LSB bits of the pixel byte.

Step 11: Advance to the next group in the pixel byte.If the secret bits are embedded goto Step 12; Else goto Step 7.

Step 12: Stop.

The following figure shows the embedding procedure:

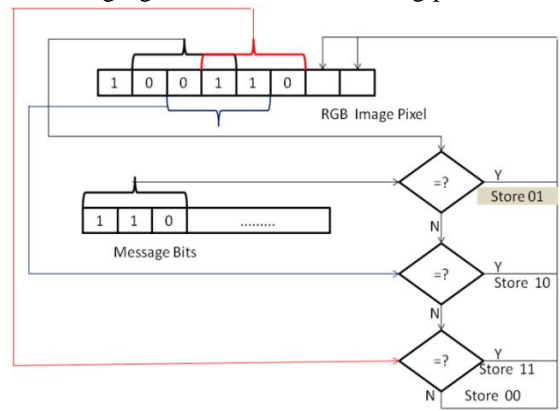

Once the stego image is received at the receiver end, the secret message can be extracted by applying the reverse of the embedding procedure. The following are the steps to be followed in the extraction procedure.

Step 1: Start

Step 2: Select a RGB cover image and the secret text to be embedded.

Step 3: Encode the cover image and hidden text into binary.

Step 4: Choose a pixel and divide it into 3 bytes(color component). Select a byte and group the bits to form 3-bit group.

Step 5: Check the value of the lower two LSBs of the component to get number n between 0-3.

Step 6: if n==0 then go to Step 8.

Step 7: The nth group in the component is considered as the group which matches the required secret bit group.

Step 8: Advance to next group in the pixel byte. If all the secret bits are extracted, go to Step 9; Else go to Step 5.

Step 9: Stop.

The length of the secret message can be embedded in the buffer i.e. a separate pixels which are meant to contain system defined information. In the receiver side, at first, these information can be extracted there by knowing the length of the secret message to be extracted.

### IV EXPREIMENTAL RESULTS AND DISCUSSION

Experimental results for both the algorithms are included in this section. For both the algorithm, standard RGB color image 'Lenna' is used to measure the performance. The text message used to embed in the image is "Steganography is called covered writing".

The Figure 1 shows the original lenna image and Figure 2 shows the stego image from the second algorithm.

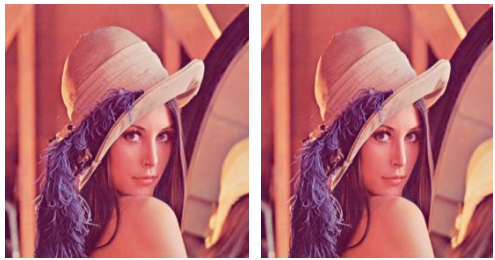

Figure1                    Figure 2

The following table shows the PSNR value of the stego image by using both the. Although distortion is not visible to naked eye, PSNR value indicates the quality of the stego image.

| Cover Image | Algorithm used | PSNR(dB) |
|---|---|---|
| Lenna.png | Propose algorithm | 75.07 |
| Lenna.png | Vikesh et al pair and matching [1] | 102.94 |

## V. CONCLUSION

The proposed algorithm and Vikesh et al Pair and Matching [1] algorithm is implemented and the PSNR values are recorded to measure the stability of the algorithm. Even though the PSNR value of proposed algorithm is lower than that of Pair and matching technique [1], the number of bits inserted is more. Three cases are considered in the output behavior of the proposed algorithm. In the best case, all the comparison results in the match; we can embed the text that is 3 times longer than that of the simple LSB algorithm. In the worst case no data could be embedded due to all mismatches which are very rare. The average case behavior is observed generally which can hold the data more than that of pair and matching technique [1] but lesser than the best case scenario.

The proposed algorithm is an attempt to store more data in the image where the security is of less concern. Even though the PSNR value is low, it can be a good foundation to build more secure algorithms.